\documentclass[lettersize,journal]{IEEEtran}
\usepackage{amsmath,amsfonts}
\usepackage{algorithmic}
\usepackage{algorithm}
\usepackage{array}
\usepackage[caption=false,font=normalsize,labelfont=sf,textfont=sf]{subfig}
\usepackage{textcomp}
\usepackage{stfloats}
\usepackage{url}
\usepackage{verbatim}
\usepackage{graphicx}
\usepackage{cite}
\newcommand{\textcc}[1]{\textcircled{\scriptsize#1}}
\usepackage{multirow}

\begin{document}

\title{Grouped Target Tracking and Seamless People Counting with a 24 GHz MIMO FMCW}

\author{Dingyang Wang,~\IEEEmembership{Member, IEEE},
Sen Yuan,~\IEEEmembership{Member, IEEE},
Alexander Yarovoy,~\IEEEmembership{Fellow, IEEE}, Francesco Fioranelli,~\IEEEmembership{Senior Member, IEEE}
\thanks{This work involved human subjects or animals in its research. Ethic approval was granted by the TU
Delft HREC.}

\thanks{The authors are with the Microwave Sensing Signals \& Systems (MS3) Group, Department of Microelectronics, Delft University of Technology, 2628 CD Delft, The Netherlands
(e-mail: D.Wang-6@tudelft.nl; S.Yuan-3@tudelft.nl; F.Fioranelli@tudelft.nl; A.Yarovoy@tudelft.nl).}
 
}

\markboth{Journal of \LaTeX\ Class Files,~Vol.~14, No.~8, August~2021}%
{Shell \MakeLowercase{\textit{et al.}}: A Sample Article Using IEEEtran.cls for IEEE Journals}


\maketitle

\begin{abstract}
The problem of radar-based tracking of groups of people moving together and counting their numbers in indoor environments is considered here.
A novel processing pipeline to track groups of people moving together and count their numbers is proposed and validated. 
The pipeline is specifically designed to deal with frequent changes of direction and stop \& go movements typical of indoor activities. 
The proposed approach combines a tracker with a classifier to count the number of grouped people; this uses both spatial features extracted from range-azimuth maps, and Doppler frequency features extracted with wavelet decomposition. Thus, the pipeline outputs over time both the location and number of people present.
The proposed approach is verified with experimental data collected with a 24 GHz Frequency Modulated Continuous Wave (FMCW) radar. It is shown that the proposed method achieves 95.59\% accuracy in counting the number of people, and a tracking metric OSPA of 0.338.
Furthermore, the performance is analyzed as a function of different relevant variables such as feature combinations and scenarios. 

\end{abstract}


\section{Introduction}
\IEEEPARstart{R}{adar-based} human tracking is a very active research field, leveraging on the advantages of using radar sensors. 
Firstly, they operate contactless and do not require additional equipment to be attached or worn by the users. 
Secondly, they are expected to be more respectful of personal privacy than vision-based sensors, and 
are insensitive to ambient light conditions or glaring. 
In combination with a compact multiple input multiple output (MIMO) Frequency Modulated Continuous Wave (FMCW) systems, these features make radar a very attractive sensor for indoor observation and tracking of humans.



In the literature, there are different approaches for indoor human tracking \cite{MultipleTargetPositioningyoo2019} leveraging on multiple domain information obtainable by radar sensors. These are typically based on the Range-Azimuth (RA) domain \cite{PeopleTrackingandCountingReferenceDesigntexasinstruments}, or the Range-Doppler (RD) domain \cite{RadarBasedRobustPeopleninos2022,NoncontactExtractionBiomechanicalwang2021,GroupedPeopleCountingren2023}, as the starting points for detecting the presence of targets.
Two processing pipelines starting from detections on the RA vs RD maps were formulated and compared in our previous work in \cite{AnalysisProcessingPipelineswang2024}. Furthermore, an additional comparison on the processing steps of data association and tracking methods was presented in \cite{QuantitativeAssessmentPeoplewang2024}. The OSPA (Optimal Sub-Pattern Assignment Metric) metric was used in that work for performance assessment.

\begin{figure}[!t]
\centering
\includegraphics[width=0.5\linewidth]{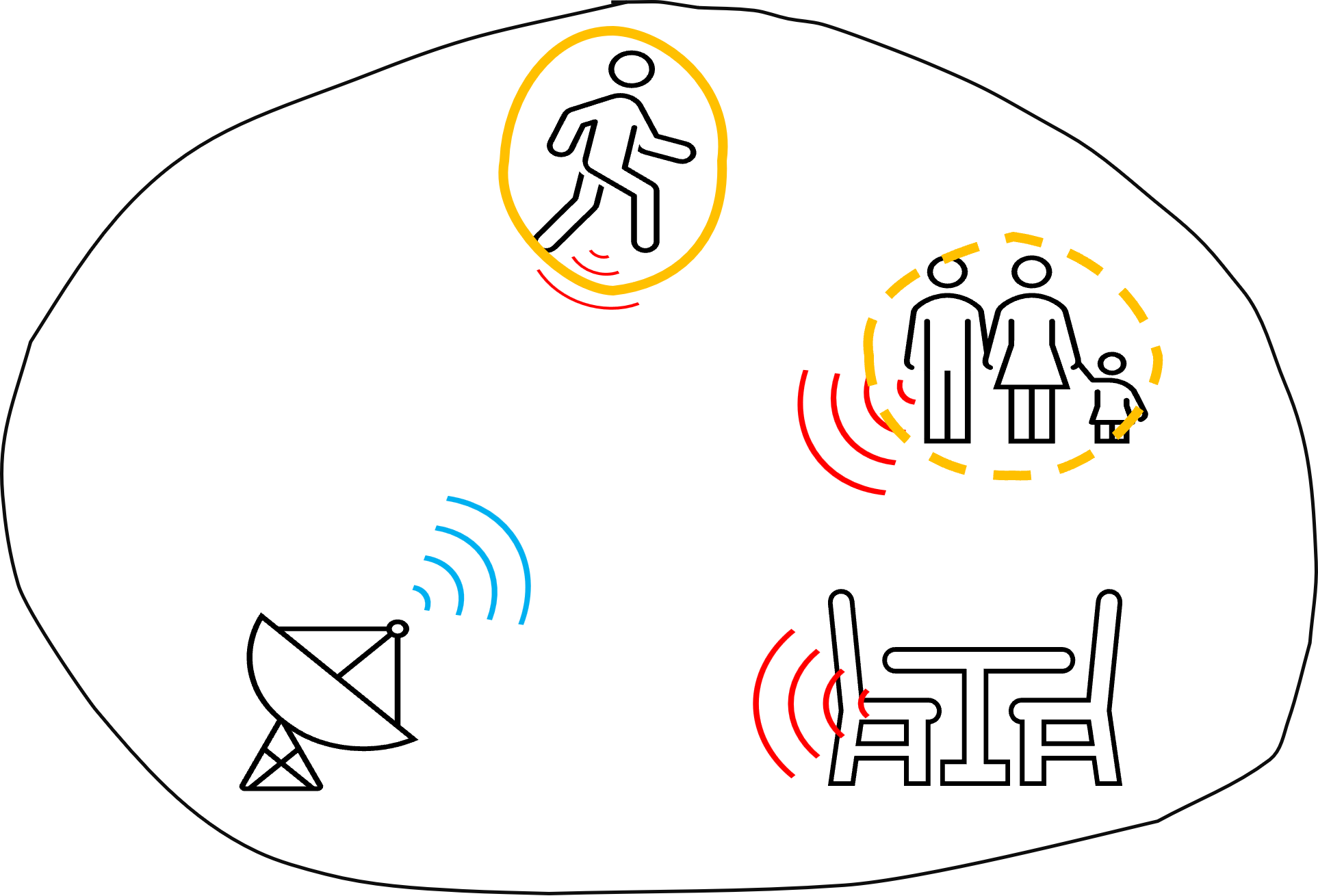}
\caption{Sketch of scenario with single target (one person) and grouped target (two or more people close by).}
\label{f.groupedTar}
\end{figure}

While promising, these initial results showed that tracking and counting the number of people moving as a group remained challenging.
A `grouped target' can be defined as a cluster of people moving together while remaining close to each other, for example shoulder by shoulder or following each other. A simple sketch is shown in Fig. \ref{f.groupedTar} to provide an example.
This behavior is very common in daily life, e.g. \, a couple of colleagues walking together in a corridor to the office, or a family with a kid crossing a road section. 
From the point of view of radar, the challenge comes from the insufficient spatial resolution to separate people in the group and the mixing of their (micro-)Doppler signatures. Additionally, in indoor environments there are very frequent changes of direction and velocity (e.g., sort of stop \& go movement pattern) due to the limited physical space for people to move and occlusion by furniture. This makes the Doppler signatures of grouped people even more confusing to analyze to estimate how many people are present in the group.


In order to address the above challenge, in this work the problem is formulated as `grouped target' tracking instead of tracking individual subjects. Simultaneously, the proposed processing pipeline combines a classifier to count the number of people present in each tracked group.
Unlike the previous initial work in \cite{AnalysisProcessingPipelineswang2024,QuantitativeAssessmentPeoplewang2024}, the proposed method takes into account the grouped nature of the targets. 
Additionally, unlike the work in \cite{GroupedPeopleCountingren2023} where performance were tested in outdoor scenarios with relatively low clutter and wide spaces generating little multipath, here a cluttered indoor space is considered. To deal with the more challenging signatures resulting from this, features based on wavelet decomposition in the Doppler domain are formulated and used to count the number of people. These outperform the cadence velocity diagram (CVD) based features used in \cite{GroupedPeopleCountingren2023}. In general, it is shown that the accuracy of the proposed approach improved by 7.12 \% compared to that method.


Summarizing, the main contributions of this work are:
\begin{itemize}
        \item The problem of tracking and counting grouped targets in an indoor environment is addressed. A dedicated processing pipeline is proposed to utilize the information from a MIMO FMCW radar combining tracking and classification. 
        \item The pipeline specifically combines tracking with extended Kalman filter and a classifier to estimate the number of people in the group area. The classifier is trained with spatial features from range-azimuth maps and Doppler frequency features derived from wavelet decomposition. 
        \item The proposed method is experimentally tested with a 24 GHz FMCW MIMO radar in a laboratory room which is an environment with heavy clutter and multi-path. The proposed method achieves 95.59\% accuracy in counting the number of people and shows an average error of 0.338 OSPA. 
\end{itemize}

The rest of the paper is organized as follows. In Section \ref{s.proposed_method}, the proposed pipeline for simultaneously tracking \& counting people is presented, along with the utilized features. In Section \ref{s.exp}, the experimental setup and details of the data collection are discussed. The performance analysis and the effect of the different variables is provided in Section \ref{s.results}. Finally, conclusions are given in Section \ref{s.conclusion}.

\section{Proposed Method}
\label{s.proposed_method}

\begin{figure*}[!t]
\centering
\subfloat[Tracking pipeline]{\label{f.algoa}
\includegraphics[height=0.3\linewidth]{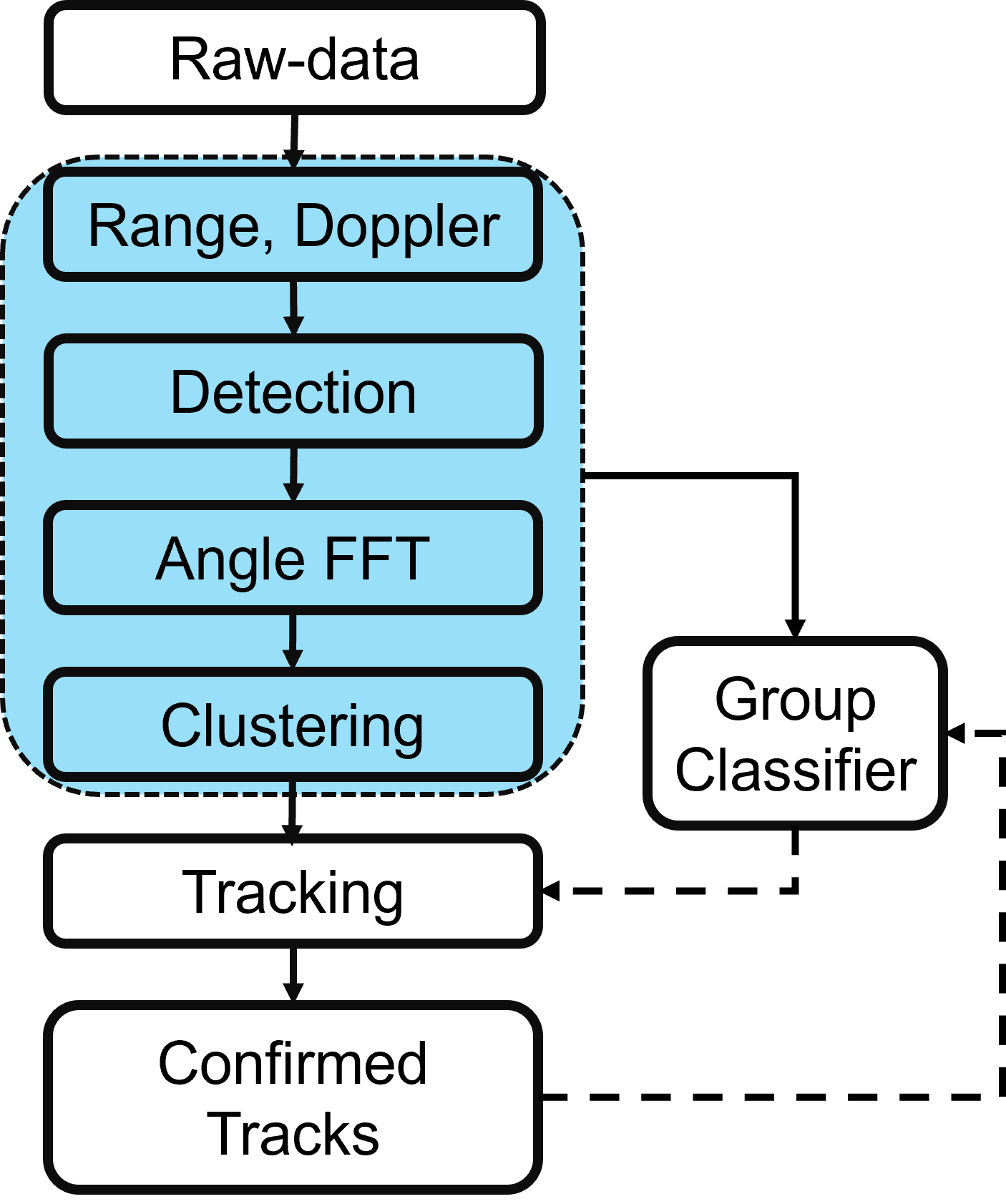}}
\;\;\;\;\;\;\;\;\;\;\;\;
\subfloat[Classifier pipeline]{\label{f.algob}
\includegraphics[height=0.3\linewidth]{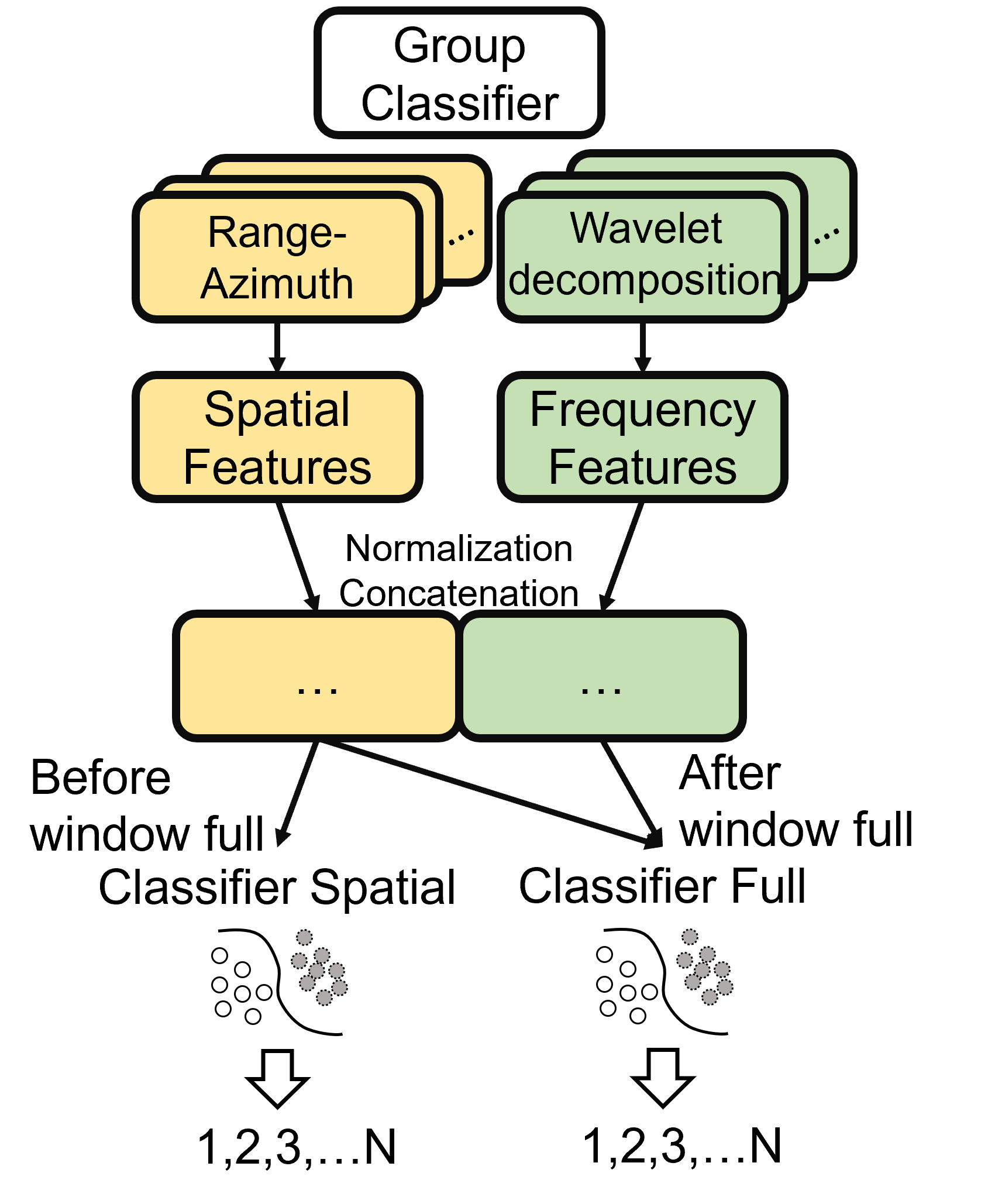}}
\caption{Pipeline of the proposed approach for grouped target tracking with integrated classifier.}
\label{f.algo}
\end{figure*}

In this section, the processing pipeline proposed is presented in two parts as shown in Fig \ref{f.algo}. The first part on the left-hand side is about the tracking method based on detections from the Range-Doppler maps. On the right-hand side, the seamless group classifier for counting the number of people is introduced in the second part of the pipeline. In order to get enough data to extract Doppler frequency features, an observation time of suitable duration is required. Before this time window is complete, the classifier only uses spatial features based on range-azimuth maps for initial counting. Additional details for the two parts of the pipeline are presented in the following sections.

\subsection{Tracking}
\label{ss.racking}

The left part of Fig. \ref{f.algo} shows essentially a conventional tracking pipeline. Detections from the Range-Doppler maps are used as the starting point of the processing, leveraging on the relatively finer Doppler resolution compared to the angular one. More details on performance differences on using detections from Range-Doppler maps rather than Range-Azimuth maps were discussed in our previous work \cite{AnalysisProcessingPipelineswang2024}. After detection, FFT (Fast Fourier Transform) is performed to estimate azimuth angles of the detected targets and radar cubes are obtained. Then, a clustering method is applied, specifically the Density-Based Spatial Clustering of Applications with Noise (DBSCAN) algorithm \cite{DensitybasedAlgorithmDiscoveringester1996} is used in this work. DBSCAN is particularly well-suited for extended targets due to its ability to handle irregularly shaped clusters and automatically identify outlier points (e.g., noise) within the data. This step can help differentiate between multiple targets and reduce false alarms when dealing with point clouds. If multiple targets are close to each other and within the resolution cell, there will be one cluster according to the DBSCAN hyperparameters. The center point of the cluster will represent the cluster position for further tracking. 

Following clustering, the data association step is used to find an optimal match between new detections and current tracks. The process steps are generally based on the Extended Kalman filter framework \cite{IntroductionKalmanFilterwelch1995} and denoted by the `Tracking' block in Fig. \ref{f.algo}. Three different data association algorithms were investigated by us earlier in \cite{QuantitativeAssessmentPeoplewang2024} resulting in a selection of the Global nearest-neighbor (GNN) \cite{ReviewStatisticalDatacox1993} for this study. The GNN method considers all possible pairings between new detections and current tracks. The weight is calculated according to a distance-based cost function. The most likely association is made from the weight. Compared with the Joint probabilities data association (JPDA) \cite{ProbabilisticDataAssociation2009a}, GNN provides a hard rather than a soft assignment, which is able to directly find the detection contribution to tracking.

\subsection{Feature extraction}
\label{s2.feature_extracting}

\begin{figure}[tb]
\centering
\includegraphics[width=0.8\linewidth]{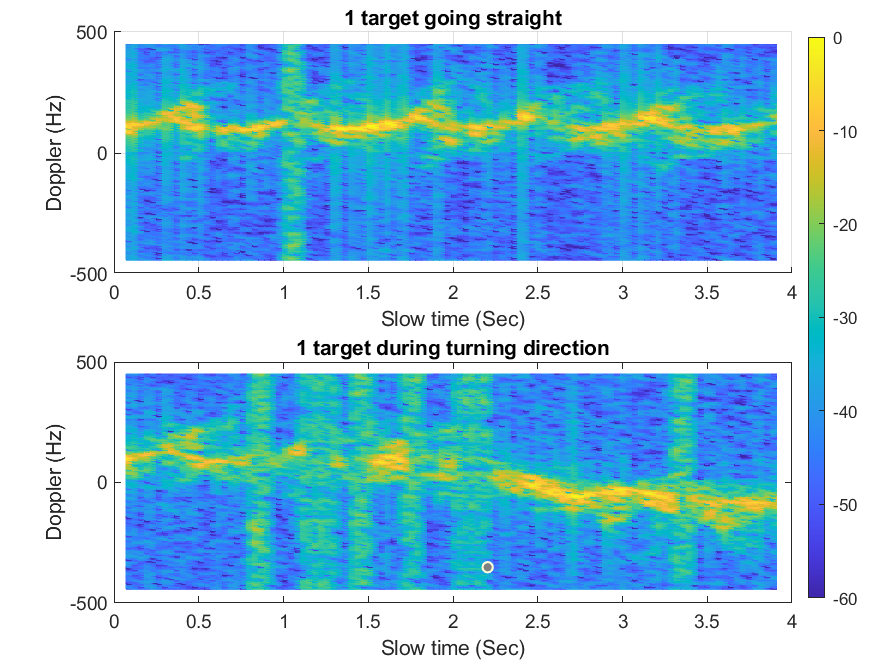}
\caption{Comparison of spectrograms between the case of `normal' walking in one direction and walking with direction changing.}
\label{f.1TDopdiff}
\end{figure}

In the previous study \cite{GroupedPeopleCountingren2023}, features are extracted from the range-azimuth maps and cadence velocity diagrams (CVD) maps. 
The CVD $S_C(\epsilon, k)$ is obtained by taking the FFT of the Doppler spectrum across the time axis, as:
\begin{equation}
        S_C(\epsilon , k) = \sum_{l=0}^{N_w-1}| \hat S(l, k) | w(l) \exp \left(-j2\pi \frac{\epsilon l}{N_w} \right)
\end{equation}
where $S(l, k)$ is the spectrogram, $w(l)$ is a window function with length equal to $N_h$. $N_w$ is the total number of windows when performing the FFT. $\epsilon = 0,1,2,..., N_{h-1}$ and $ k = 0,1,2,..., N_{w-1}$ denote the cadence frequency index and Doppler index respectively.
In order to estimate the proper frequency from the CVD map with sufficiently fine resolution, a longer observation window is required. If the observation of people is performed in an outdoor environment, the long observation window may not be a problem as typical people move along longer and more regular trajectories. However, when considering an indoor environment with walls and furniture in normal-sized rooms, due to the limited moving space, people can change their direction or stop and go again within the ideal observation window time.
As shown in Fig. \ref{f.1TDopdiff}, the Doppler pattern changes completely due to the direction turning within the observation window. Therefore, in these cases frequency features directly extracted from the CVD may not be reliable. 

To extract reliable frequency features related to the Doppler domain (and therefore target velocity), wavelet-based methods are a good approach to decompose the original signal into different levels. Furthermore, it is important to see how each decomposition level can contribute to the final accuracy, so that levels containing little information for the problem at hand can be discarded and not be used by the classifiers. 
In this work, the Maximal overlap discrete wavelet transform (MODWT) \cite{MaximalOverlapDiscretexiao2020,MaximalOverlapDiscretequilty2021,ChangeDetectionTimealarcon-aquino2009} is used for this purpose. 
First, the time series data to be analyzed is denoted as $\mathbf{X} = \{X_t\}$ where $t=0,...,T-1$, $T$ is the total number of frames for the observation window, and $X_t$ represents the range bin values corresponding to the target's location at frame $t$. Note that $X_t$ is a vector, as each frame consists of multiple chirps. The $j$th level wavelet and scaling filter are denoted as $\{\tilde h_{j,l}\}$ and $\{\tilde g_{j,l}\}$, respectively. $l$ is the index of filter coefficients. The scaling and wavelet coefficient can then be expressed as follows:

\begin{equation}
        \tilde V_{j,t} = \sum_{l=0}^{T-1} \tilde g_{j,l} X_{t-l \; mod\; T}
\end{equation}

\begin{equation}
        \tilde W_{j,t} = \sum_{l=0}^{T-1} \tilde h_{j,l} X_{t-l \; mod\; T}
\end{equation}
where $j=1,2,...,J$ is the level of the wavelet decomposition, in this work specifically $J=4$.

Compared to the normal wavelet method mentioned in \cite{ChangeDetectionTimealarcon-aquino2009}, the MODWT approach avoids down-sampling and prevents data size to be too low with the increasing number of levels. Additionally, the MODWT preserves time information alignment and provides multi-resolution frequency analysis.
For statistical feature extraction and subsequent usage into classifiers, having the same data size for each level is important for consistency. 

\begin{figure}[tb]
\centering
\includegraphics[width=0.9\linewidth]{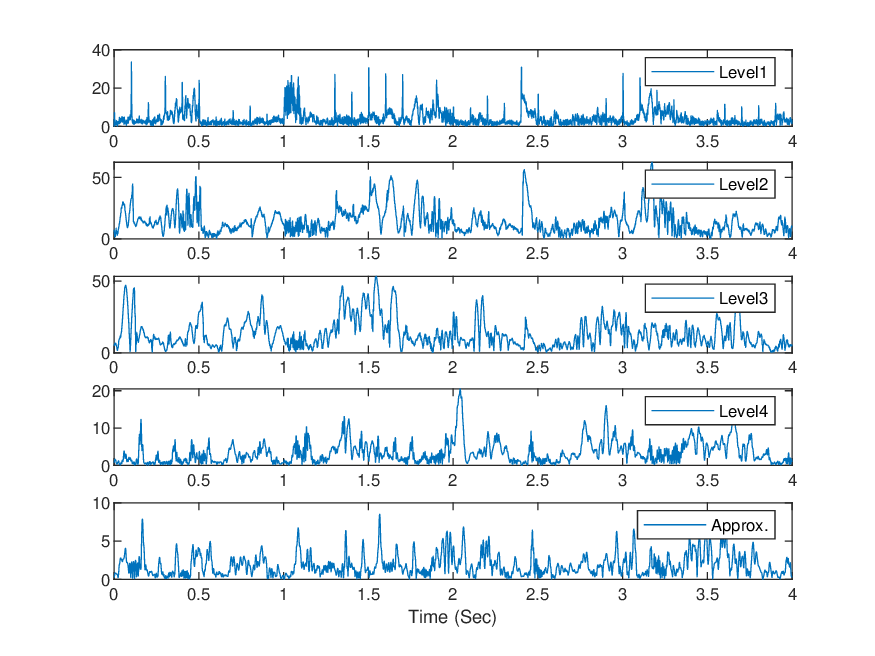}
\caption{Example of MODWT decomposition from data for the case of 1 target walking along a straight trajectory.}
\label{f.1TmofwtL4}
\end{figure}

In order to implement this method and extract the relevant MODWT features, a tracking trajectory at time $t$ and tracking ID is needed and denoted as $Trac_{t}^{ID}$. For a given ID, the range bin data $X_{t}^{ID}$ 
and angle information are stored in a buffer to extract the features.
The spatial features are those extracted from the range-azimuth map and are specifically the width occupied by the target in the azimuth axis, the length occupied in the range axis, the mean value of angle bins, the median value of angle bins, the variance of angle bins, the number of different angle bins occupied, the mean value of angle profile, the median value of angle profile, the variance of angle profile, and the number of pixels occupied in the range-azimuth map. The spatial features are extracted from each window frame and averaged along the time axis to smooth their values. 
The frequency features are extracted from the 4 levels of MODWT. 
There are eight statistical features for each level including the variance, standard deviation, mean value, median value, root mean square value, skewness value, kurtosis value, and entropy value.

An example of MODWT decomposition is shown in Fig. \ref{f.1TmofwtL4} with its time axis aligned to the upper sub-plot of Fig. \ref{f.1TDopdiff}.
In the case of missed detections while performing the tracking process, there will not be related data stored in the buffer for a few time bins. In these cases, the mean value of the remaining existing data will be used to fill the missing samples in that observation window. 
With the chosen 4 layers for decomposition, the resulting frequency bands are level 1 from 225 to 450 Hz, level 2 from 109 to 233 Hz, level 3 from 54 to 116 Hz, level 4 from 27 to 58 Hz, and `Approx' level from 0 to 28 Hz.
These are also listed in Table \ref{t.accmodwtLevel} later in the paper. 

It should be noted that artifacts such as those in the Doppler spectrum around 1s in Fig. \ref{f.1TDopdiff} are filtered and categorized as belonging to the high frequency level in Fig. \ref{f.1TmofwtL4} after MODWT decomposition. In the later results section, the effect of selecting features from different frequency bands will be analyzed. This ability to separate spectral artifacts from useful signals is one of the advantages of using wavelet-based method for extracting features.

\begin{figure}[tb]
\centering
\subfloat[Comparison of number of people VS. angle bin width]{\label{f.feaNo1}
\includegraphics[width=0.8\linewidth]{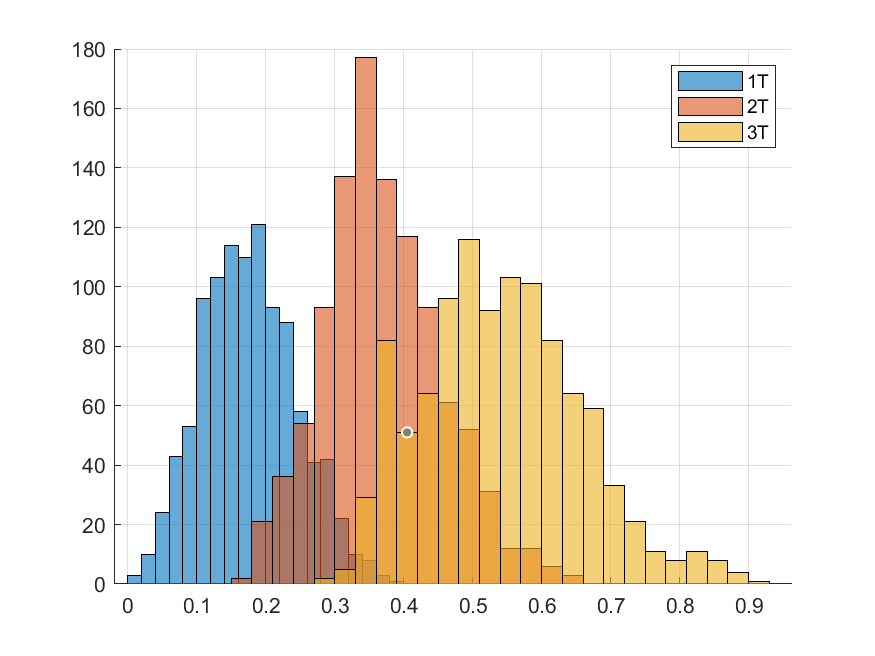}}
\,
\subfloat[Comparison of number of people VS. mean value of Level 3]{\label{f.feaNo29}
\includegraphics[width=0.8\linewidth]{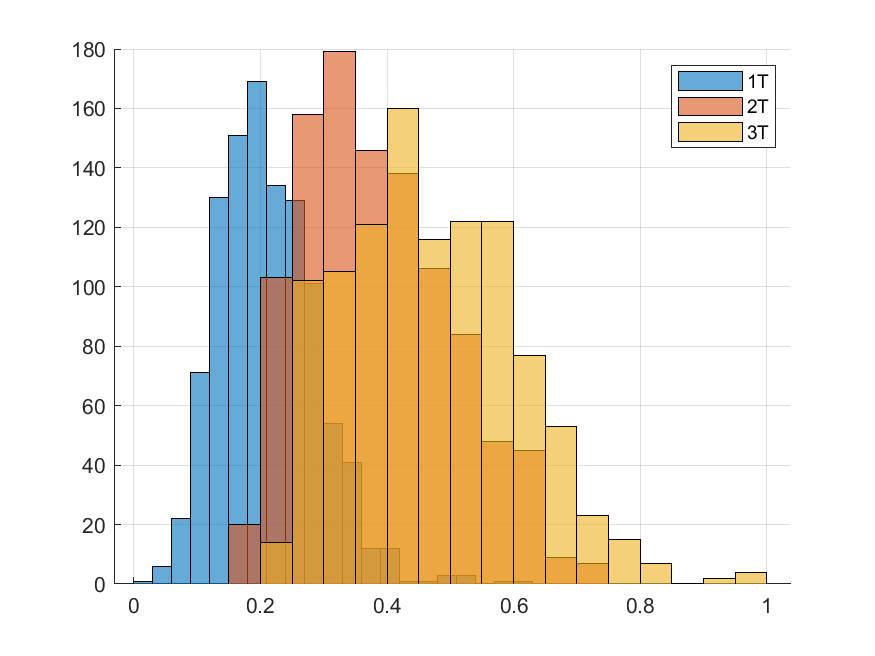}}
\caption{Histograms with distribution of feature values for the cases of 1, 2, and 3 people.}
\label{f.feaComp}
\end{figure}

To better understand how the number of people affects the radar cross-section (RCS) and frequency characteristics, refer to Fig. \ref{f.feaComp}, which illustrates representative features from both the spatial and frequency domains. A clearer separation between each target's distributions enhances the classification accuracy.
The width of the intensity in the range-azimuth domain, measured in angle bins, varies with the overall RCS of the group as the number of people changes. Additionally, the amplitude at Level 3 increases as the number of people increases. Level 3 has a frequency range spanning from 54 to 116 Hz, which corresponds to the band of Doppler frequencies mostly occupied by human walking activities.

\subsection{Classifier}
\label{ss.classifier}


The classifier assigns a label to a given set of input features. In this study, the assigned label corresponds to the number of individuals in a group.
Given the relatively limited size of the available datasets in this context, classification methods that do not rely on neural networks (NNs) are considered. Various classification algorithms exist, each exhibiting different performance characteristics depending on the extracted features and the specific classification task.
To present comprehensive results, four common non-NN classification algorithms are selected and compared. They are namely: 1) k-nearest Neighbors (KNN) \cite{KNNModelBasedApproachguo2003}, 2) Naïve Bayes \cite{EmpiricalStudyNaiverisha}, 3) Support Vector Machine (SVM) \cite{SupportVectorMachineshearst1998} and 4) Random Forest \cite{RandomForestRemotebelgiu2016}.
The SVM is a powerful and versatile supervised learning algorithm used for classification tasks. SVM aims to find the optimal hyperplane that maximizes the margin between different classes. The margin is the distance between the hyperplane and the nearest data points from each class, known as support vectors.
The SVM is calculated by minimizing:
    \begin{equation}
        \left[\frac{1}{n} \sum_{i=1}^{n} \max\left(0, 1-N_i(w^T x_i -b)\right) \right]+\lambda \| w \|^2,
    \end{equation}
where $\max\left(0, 1-N_i(w^T x_i -b)\right)$ is the Hinge loss used for training the SVM classifiers, $w$ and $b$ are used to calculate the estimated labels, $N_i$ is the $i$-th data points. $\lambda \| w \|^2$ is a term introduced to limit the margin size and make $x_i$ assigned on the correct side.
SVMs can be rather sensitive to the choice of the kernel and its parameters. In this case, the SVM classifier was created by using a Gaussian kernel which appeared to be the most suitable choice for this work. 
The data is divided into 70\% for training and 30\% for testing. 

After obtaining the output of the classifier for each time/observation window, there is a median filter applied to make the output smoother and remove temporary prediction errors. 

\subsection{Seamless counting}
\label{ss.smlesCounting}
To extract suitable Doppler frequency features using the MODWT approach, a longer observation window is required. However, this window exceeds the time window used in the tracking process for data association and track confirmation. The difference in processing speeds between these two stages creates a time gap between trajectory confirmation and the availability of sufficient data for accurate Doppler frequency feature extraction.
In contrast, spatial features derived from range-azimuth maps are available immediately for each frame. These features can be used for an initial counting estimation, ensuring continuous, `seamless' results while more data are being acquired.

\begin{figure}[tb]
\centering
\includegraphics[width=0.9\linewidth]{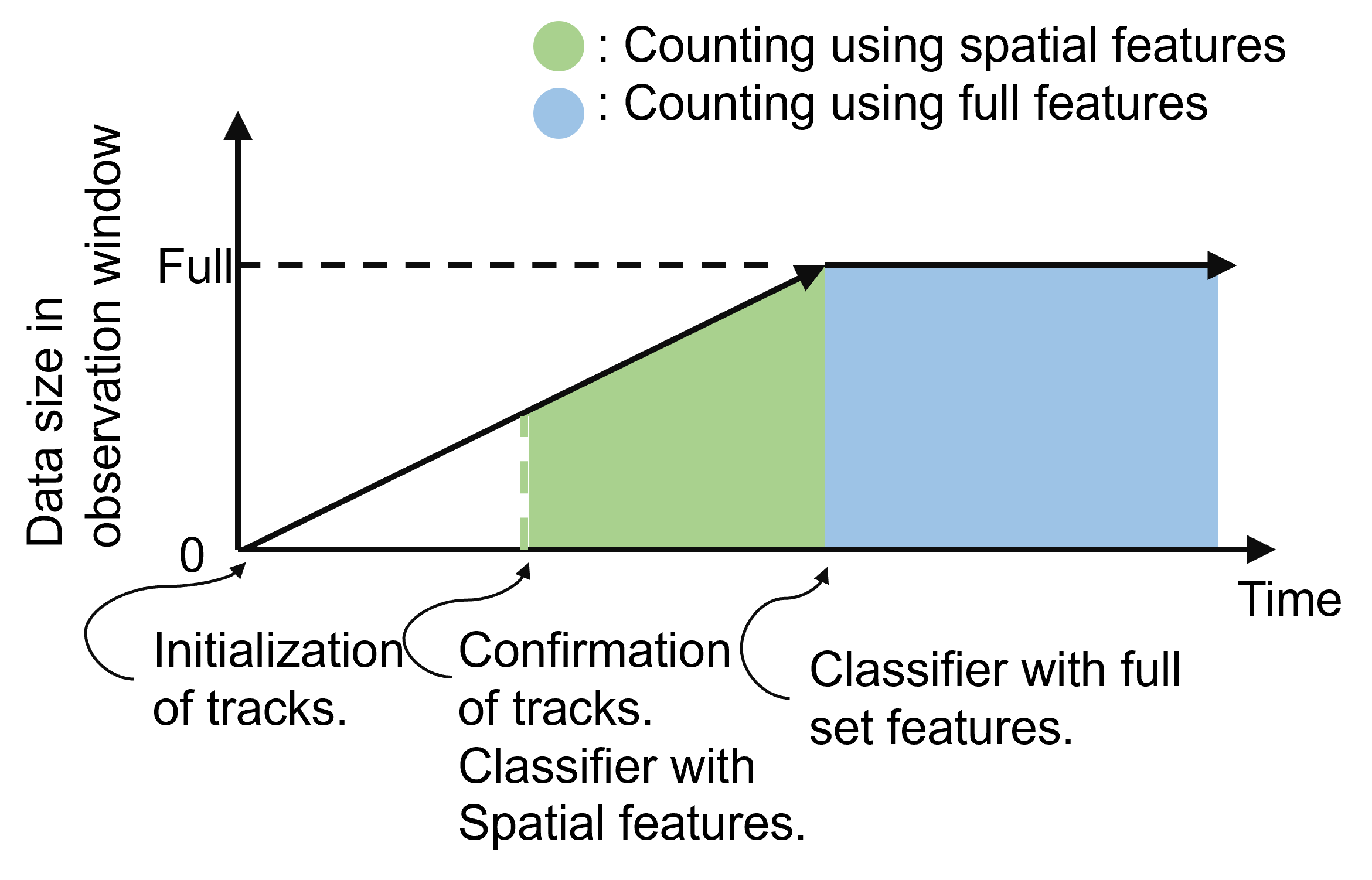}
\caption{Diagram of the proposed tracking approach with integrated seamless counting over time. There is an interval while the classifier can only utilized a subset of features while the data are still being collected.}
\label{f.seamlCount}
\end{figure}

Fig. \ref{f.seamlCount} presents an illustration of the seamless counting process. The green zone represents the time interval during which only spatial features are available; however, the data are insufficient to reliably compute Doppler frequency-based features. The blue zone indicates the period when the classifier can perform people counting using the full set of features.

In general, once a track is initialized, the amount of data available for extracting spatial features gradually increases. The counting classifier, based on spatial features, becomes active upon track confirmation and continues operating until the observation time window is complete. After this point, the classifier incorporates both spatial and frequency features for improved accuracy.

For a summarizing overview of the proposed tracking combined with the group classifier, it is noted that the method processes the current data input from the Range-Azimuth-Doppler domains along with previously extracted features from confirmed tracks with the same tracking ID. After making a prediction, the classifier passes the predicted count of the number of people to the tracking block in a sort of feedback connection, which in turns helps managing the number of active tracks.

\section{Experimental Setup}
\label{s.exp}
This section presents the experimental setup, including the radar and reference sensor. Data was collected across six scenarios to validate the proposed method. In addition, the metrics used to quantify the results is introduced.

\subsection{Radar sensor}
A commercial 24GHz FMCW radar (by Joby Austria, former INRAS) with a relatively low bandwidth of 250MHz is used to evaluate the performance. The relatively small bandwidth results in a range resolution of approximately 60 cm, which indicates that the target occupies a large area in the range profile, while also making it difficult to isolate individual body parts. The detailed parameters used are listed in Table \ref{t.fmcwpara}.
The radar is equipped with 15 virtual channels in azimuth which provide a fine angular resolution (7.63 $^{\circ}$). Furthermore, an analysis of the performance when using a different number of receiving channels is possible.

\begin{table}[h]
\centering
\caption{JOBY (former INRAS) FMCW radar parameters}
\label{t.fmcwpara}
\begin{tabular}{ll}
\hline
FMCW radar model                        & RadarBook2 (RBK2)     \\
Operating frequency                     & 24 GHz         \\
Sweep bandwidth                         & 250 MHz        \\ 
ADC sampling rate                       & 120 ksps        \\
ADC samples per chirp                            & 56              \\
Up chirp duration                       & 467 $\mu$s          \\
Chirp repetition interval               & 483 $\mu$s          \\
Number of chirps in a frame             & 90 \\
Slow-time sampling frequency            & 10 Hz         \\
Number of TX \& RX channels             & 2 x 8 \\
Antenna horizontal 3 dB beamwidth       & 76.5$^{\circ}$ \\
\hline
\end{tabular}
\end{table}

\subsection{Environment for data collection}
\begin{figure}[!htb]
\centering
\subfloat[Experimental room]{\label{f.expEnv1}
\includegraphics[width=0.45\linewidth]{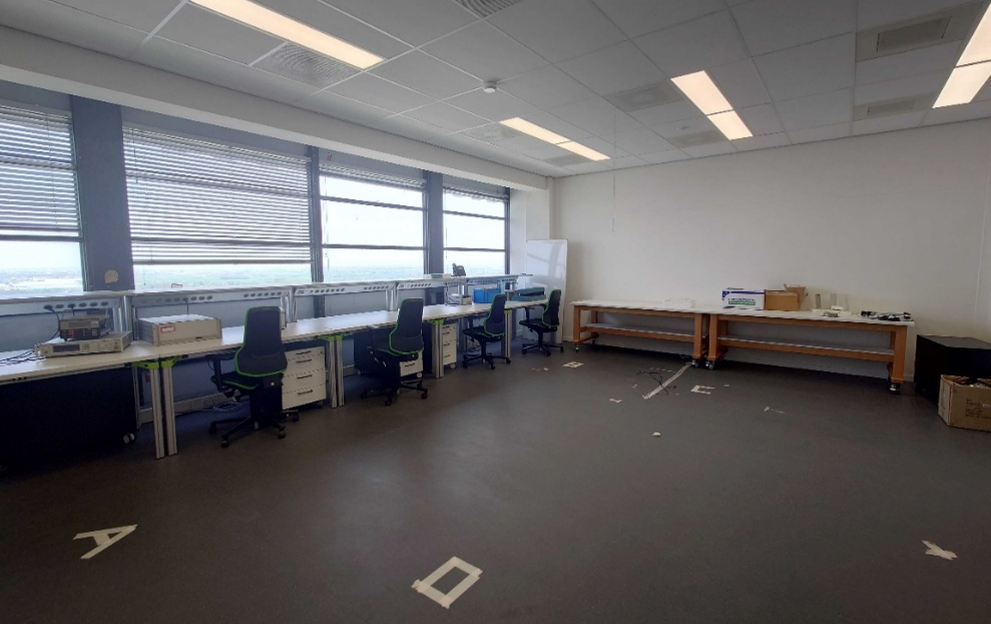}}
\,
\subfloat[Setup of INRAS radar and Azure Kinect DK camera]{\label{f.expEnv2}
\includegraphics[width=0.45\linewidth]{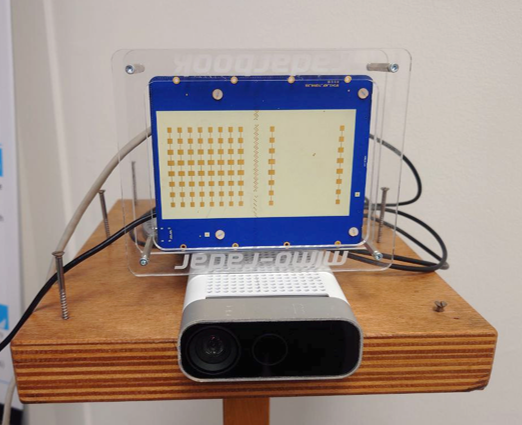}}
\caption{Experimental environment for data collection in the MS3 Radar Laboratory at TU Delft.}
\label{f.expEnv}
\end{figure}

For the purpose of this research, a data set was specifically collected with 5 individuals, and up to 3 persons were moving simultaneously in the laboratory room of the MS3 group at TU Delft. Essentially, the number of people simultaneously moving in the environment were 1, 2, or 3.
To emulate the cluttered environment of normal office space, pieces of furniture such as tables, chairs, and cabinets were placed in the environment, and a metallic curtain was also present at the window, which contributed to the relatively high level of multi-path recorded. 
As illustrated in Fig. \ref{f.expEnv1}, a normal laboratory room with open space in the center is presented. 
The radar was placed at around 1.3-meter height in the corner of the room, with its line of sight pointing along the diagonal of the room to get wide coverage, as shown in Fig. \ref{f.expEnv2}.
In total, 6 different movement scenarios are performed and collected. These activities are listed in Table \ref{t.activities}, where it is shown that they include a mixture of different walking patterns with variable distances between the people in case of groups of them being present.
Most of the activities are recorded for around 20 minutes. This leads to having around 51000 frames of data collected and used for validation.

\begin{table}[!htb]
\centering
\caption{Summary of the 6 different movement scenarios collected for testing the proposed method.}
\label{t.activities}
\begin{tabular}{ccc}
\hline
\#NO. & Movement Scenarios                       & Duration (min) \\
\hline
\textcc{1}     & 1 Target walking forward and backward   & 20             \\
\textcc{2}     & 1 Target randomly walking                 & 10             \\
\textcc{3}     & 2 Targets walking forward and backward   & 20             \\
\textcc{4}     & 2 Targets following each other while walking  & 5              \\
\textcc{5}     & 2 Targets randomly walking                 & 10             \\
\textcc{6}     & 3 Targets walking forward and backward   & 20             \\
\hline
\end{tabular}
\end{table}

\subsection{Reference sensor}
To assess and compare different tracking algorithms in the proposed pipeline, a reference sensor is needed to act as ground truth. 
To quantitatively analyze the tracking performance, an auxiliary RGBD camera (Azure Kinect DK \cite{AzureKinectDK}) was used to collect such ground truth data at the same time as the radar measurements. The reference method used for positioning analysis is the one in \cite{QuantitativeAssessmentPeoplewang2024}.
The RGB data was processed by \cite{Detectron2yuxinwuandalexanderkirillovandfranciscomassaand2025} for human detection, with the output being a bounding box. Its center is considered as the human position. Then the bounding box position is mapped to a depth map to obtain the 2D positioning of the human.
Notably, due to the maximum range of the camera being limited to approximately 6 meters, tracking ID switches can happen in the ground truth if people exceed that range during measurements.

\subsection{Performance Metrics}
\label{s3.metric}
For a single target scenario, assessing distance errors in tracking (e.g., median error, mean absolute error, and root mean square error) is sufficient in this context. 
To evaluate the accuracy of multi-target tracking systems, the Optimal Sub-pattern Assignment (OSPA) \cite{ConsistentMetricPerformanceschuhmacher2008} metric is considered. 
From the equation below, the OSPA metric has two components: one is the distance error (related to localization performance), and the other one is the cardinality error (related to the number of people recognized to be in the room).
\begin{equation}
OSPA= \left( d_{loc}^{p} + d_{card}^{p} \right)^{1/p}
\end{equation}
where $d_{loc} = \{\frac{1}{n} \sum_{i=1}^{m} d_c^p(x_i,y(i))\}^{1/p}$ is the localization error, and $p$ denotes the order, which is set to 2 in this case, with $x(i)$ from a list of ground truth data, and $y(i)$ from a list of tracks from the same timestamp which is assigned to $x(i)$. 
In the formulation $d_{c}(x,y) = min \{d_{b}(x,y), c \}$, $c$ is the cutoff-based distance and is set to 1 meter in this analysis. 
$d_{card} = \{ \frac{(n+q)-m}{(n+q)}c^p \}^{1/p}$ is the cardinality error component, where $n$ is the number of considered tracks, $m$ is the number of ground truth tracks.
Since the OSPA is the final metric chosen to compare the overall performance, it also includes the difference $q$ between the classifier predictions and the true number of targets present in the scene. 

\section{Results}
\label{s.results}


This section presents the results for the proposed method by comparing the four previously mentioned classifiers, and assessing the effect of using different subsets of the considered features and levels of the MODWT decomposition.
Additionally the overall performance for tracking, localization, and estimation of number of people via the OSPA metric are discussed. The proposed method is compared to CVD-based counting, as well as conventional tracking methods.

\subsection{Analysis of classification algorithm}
Results for the four considered statistical classifiers are compared. The input features are both spatial and Doppler frequency features. The accuracy of each classifier is listed in Table \ref{t.accSVMVSetc}. From this, the SVM provides the best performance among the 4 classifiers. The KNN shows better performance than the random forest method, while the Naïve Bayes method performs the worst. The potential reason is the redundancy of information between each MODWT level against its assumption on independence of features. 
The SVM used here is with a Gaussian kernel and kernel scale of 4.5. The smaller kernel scale is empirically selected to generate complex decision boundary, but might eventually cause overfitting.

\begin{table}[h]
\centering
\caption{Accuracy comparison of different classification methods. (\textit{Spatial}: features extracted from range-azimuth map; \textit{frequency}: features extracted from MODWT method)}
\label{t.accSVMVSetc}
\begin{tabular}{ccc}
\hline
Input features    & Classifier methods & Accuracy (\%) \\ \hline
Spatial + frequency & KNN                & 97.6         \\
Spatial + frequency & Naïve Bayes        & 79.9         \\
Spatial + frequency & SVM                & 99.1         \\
Spatial + frequency & Random Forest      & 93.0        \\ \hline
\end{tabular}
\end{table}

\subsection{Analysis of spatial \& frequency domain features}
In this section, the accuracy of the proposed method with a combination of spatial and frequency domain features is analyzed. By using all features together, the highest accuracy is obtained. However, it is noted that by using only features from one domain in isolation does not reduce too much the performance, with the spatial features alone offering better results than the frequency ones.
Moreover, to reduce the overall feature vector size and its redundancy, principal component analysis (PCA) is performed. The PCA reduces the dimension of the feature vector and helps prevent the model from overfitting. With 80 \% of the information retained in the PCA, the accuracy is 95.7\% as shown in Table \ref{t.accSpaVSFreq}; this is only 3.5\% lower than the original results with the full feature vector.

For comparison, the CVD-based counting method used in \cite{GroupedPeopleCountingren2023} is also implemented. As mentioned in Section \ref{s2.feature_extracting}, the CVD method could not extract suitable features during the targets' direction changes, which happen frequently in an indoor environment. The accuracy results are shown in the right column of Table \ref{t.accSpaVSFreq}. 
In particular, while the accuracy using all combined features is only slightly lower than when using MODWT features, the performance significantly declines when relying only on CVD-based frequency domain features. Furthermore, frequency features do not contribute to the accuracy in this case. This trend is also evident in the PCA (80\%) scenario, where accuracy decreases by approximately 8\%, a larger drop compared to the proposed method.


\begin{table}[h]
\centering
\caption{Accuracy comparison of the combination of spatial and frequency features. }
\label{t.accSpaVSFreq}
\begin{tabular}{ccc}
\hline
Input Features &
  \begin{tabular}[c]{@{}c@{}}Proposed\\ Accuracy (\%)\end{tabular} &
  \begin{tabular}[c]{@{}c@{}}CVD based counting \cite{GroupedPeopleCountingren2023}\\ Accuracy (\%)\end{tabular} \\ \hline
Spatial + frequency  & 99.23 & 94.69 \\
Spatial              & 94.80  & 94.77 \\
Frequency            & 92.2  & 64.01 \\
PCA (80\% principal) & 95.7  & 86.4  \\ \hline
\end{tabular}
\end{table}

\subsection{Analysis of different MODWT level features}

Unlike other wavelet approaches for signal decomposition, with the MODWT approach the same data size remains for each level. 
It is also observed by looking at the Doppler spectrum shown in Fig. \ref{f.1TDopdiff}, that the most significant components of the human signature correspond to level 2 (109 $\sim$ 233 Hz) and level 3 (54 $\sim$ 116 Hz) of the MODWT decomposition. The higher frequency band corresponds to regions with significant Doppler activity which contains less information in this work.
This is reflected in the accuracy values shown in Table \ref{t.accmodwtLevel} to use only a single MODWT level isolated as input features, where Level 2 achieves the highest accuracy.
In contrast, Level 1 exhibits lower accuracy due to containing less relevant information.
As shown in the right row of Table \ref{t.accmodwtLevel}, the highest accuracy is obtained when spatial features are combined with Level 3. This suggests that the low-frequency band likely captures critical information related to changes in movement direction, making it more effective than the high-frequency band.


\begin{table}[h]
\centering
\caption{Accuracy comparison of the combination of different levels of MODWT features. \textit{Approx} denotes the lowest frequency from 0 to 28Hz. }
\label{t.accmodwtLevel}
\begin{tabular}{cccc}
\hline
\begin{tabular}[c]{@{}c@{}}MODWT\\ Level\end{tabular} &
  \begin{tabular}[c]{@{}c@{}}Frequency \\ (Hz)\end{tabular} &
  \begin{tabular}[c]{@{}c@{}}ACC of \\ frequency features (\%)\end{tabular} &
  \begin{tabular}[c]{@{}c@{}}ACC of Spatial\\  + Frequency (\%)\end{tabular} \\ \hline
Approx & 0-28    & 87.44 & 99.05 \\
level4 & 27-58   & 79.77 & 99.19 \\
level3 & 54-116  & 91.7  & 99.46 \\
level2 & 109-233 & 92.99 & 97.88 \\
level1 & 225-450 & 89.85 & 98.51 \\ \hline
\end{tabular}
\end{table}

\subsection{Analysis of overall tracking \& counting performance}

In terms of overall people tracking and counting performance, the modified OSPA metric mentioned in Section \ref{s3.metric} can be used for quantitative analysis. This includes both position/localization error as well as a cardinality error which accounts also for possible misclassifications on the number of people by the classifiers. Examples of OSPA based performance analysis follow.

\begin{figure}[!htb]
\centering
\subfloat[Classification results before the median filter and after median filter (BM: before median filter, AM: after median filter)]{\label{f.3TFBpred}
\includegraphics[width=0.95\linewidth]{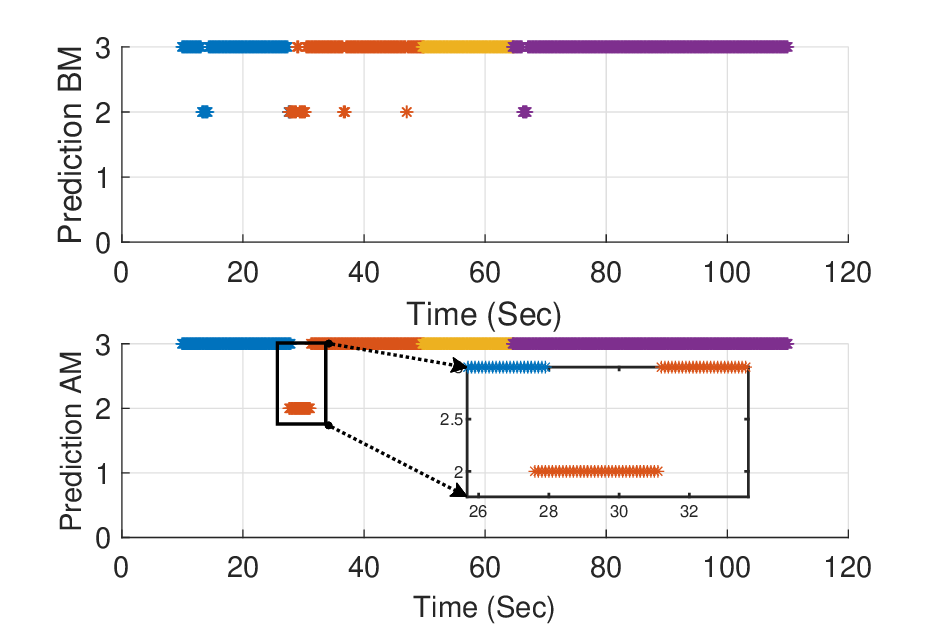}}
\,
\subfloat[OSPA results over time]{\label{f.3TFBOSPAprop}
\includegraphics[width=0.95\linewidth]{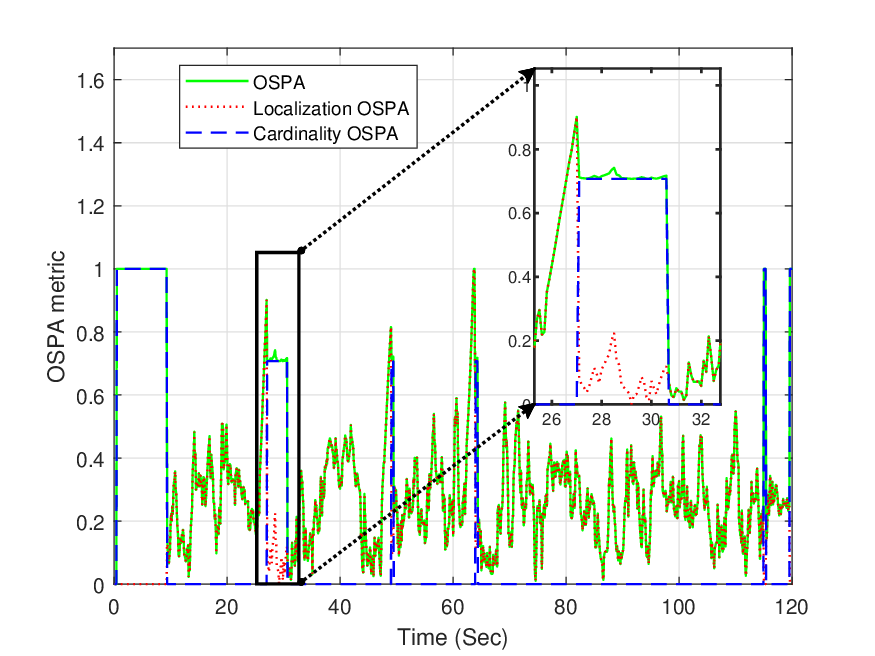}}

\caption{Analysis of results for the case of 3 people walking forward and backward. (a) number of counted people; (b) OSPA metric.}
\label{f.res3T}
\end{figure}

First, the prediction results given by the classifier for an example of 3 target scenarios are shown in Fig. \ref{f.3TFBpred}. The trajectories can be interrupted due to missed detections during tracking. The different tracking IDs are distinguished by colors. The upper plot in the figure is computed before applying the median filter. The lower plot shows the prediction after the application of the median filter with 25 samples. 
This helps remove temporarily wrong predictions and improve the final accuracy. However, the median filter also has a limitation: when the wrong predictions are consistent for more than half the window length, the error is propagated for a long time till the true value accumulates back for a long time.
This drawback can be observed from the lower plot of Fig. \ref{f.3TFBpred} around 30 seconds. This can also be seen as the limitation of statistical classifiers when applied to time-series data. The features are extracted by shifting a window considering a single block of data; however, the model does not have a feedback loop from the previous predictions to the current frame.

Then, the overall OSPA plot over time is shown in Fig. \ref{f.3TFBOSPAprop} for the same example. At the beginning of the plot, there is no confirmed trajectory. Hence, the OSPA raises its value up to 1, which is the cut-off distance defined in Section \ref{s3.metric}. Afterwards, the mean value of the OSPA remains around a mean value of 0.3. It should be noted that there is an increase in cardinality error component at about 30 seconds, which corresponds to the prediction errors of the classifier, as discussed in the previous paragraph.



An overview of detailed performance metrics for each scenario previously listed in Table \ref{t.activities} is presented in Table \ref{tab_acc}. The proposed pipeline is compared with a CVD-based counting method and a conventional tracking method (without classifier).
Specifically, the presented metrics include the classifier accuracy for counting before (BM) and after (AM) the median filter and the overall OSPA. An average performance across all scenarios is also presented for each method.
The average accuracy of the proposed method is 94.65\% and improved to 95.59\% with the median filter to eliminate temporary prediction error. 
Moreover, the counting accuracy is around 7\% higher than the CVD-based counting method. 
In terms of OSPA performance, the proposed method is 0.027 better than CVD-based counting. 
As the tracking approach is identical in both these two cases, the performance is improved by the contribution of higher classification accuracy and seamless counting. On the other hand, the conventional tracking method is not able to generate a trajectory for each target, as it cannot use the additional information on the number of people from the classifier. This reflects into higher error in the OSPA.

In most cases, the median filter helps improve the accuracy, but in the case of 2 targets following each other \textcc4 (refer to the 4th scenario from Table \ref{t.activities}) and random walking scenario \textcc5, the accuracy after the median filter is slightly less than before. The reason is that the median filter is not able to correct instances of continuous, consistent errors.
For the single target scenario, the proposed method and CVD-based counting show a bit higher OSPA than conventional tracking due to the effect of the counting error. This error is still in an acceptable range. 
Overall, it can be seen that the performance is much improved compared to the conventional tracking method.




\begin{table*}[]
\caption{Summary of performance metrics for tracking with classifier. (BM: before median filter, AM: after median filter)}
\label{tab_acc}
\centering
\begin{tabular}{cccccccc}
\hline
 \multirow{2}{*}{\textbf{Scenarios}} &
  \multicolumn{3}{c}{\textbf{Proposed}} &
  \multicolumn{3}{c}{\begin{tabular}[c]{@{}c@{}}\textbf{CVD based} \\ \textbf{counting\cite{GroupedPeopleCountingren2023}}\end{tabular}} &
  \begin{tabular}[c]{@{}c@{}}\textbf{Conventional}\\ \textbf{(tracking)}\end{tabular} \\ \cline{2-8} 
 &
  \begin{tabular}[c]{@{}c@{}}Accuracy \\ (\%) BM\end{tabular} &
  \begin{tabular}[c]{@{}c@{}}Accuracy \\ (\%) AM\end{tabular} &
  OSPA &
  \begin{tabular}[c]{@{}c@{}}Accuracy \\ (\%) BM\end{tabular} &
  \begin{tabular}[c]{@{}c@{}}Accuracy \\ (\%) AM\end{tabular} &
  OSPA &
  OSPA \\ \hline
\textcc1 1 target walking             & 94.89 & 96.22 & 0.27  & 91    & 95.11 & 0.28  & 0.24  \\
\textcc2 1 target random walking           & 85.78 & 90    & 0.34  & 89.11 & 92.56 & 0.33  & 0.3  \\
\textcc3 2 targets walking             & 99.67 & 100   & 0.21  & 88.67 & 93.22 & 0.25  & 0.73  \\
\textcc4 2 targets following            & 98.33 & 97.33 & 0.51  & 80.67 & 84.56 & 0.54  & 0.62  \\
\textcc5 2 targets random walking          & 92.67 & 91.89 & 0.4  & 73    & 78.56 & 0.45  & 0.73  \\
\textcc6 3 targets walking            & 96.56 & 97.33 & 0.3   & 76.44 & 86.78 & 0.34  & 0.85  \\ \hline
\textbf{Average}                   & 94.65 & 95.59 & 0.338 & 83.14 & 88.47 & 0.365 & 0.578 \\ \hline
\end{tabular}
\end{table*}

\subsection{Additional discussion on results}
\label{s.discussion}

The problem of counting the number of people has been approached as a classification task, as the target variable to be predicted is inherently discrete. Although this could be framed as a regression task, predicting a continuous value for the number of people would be less practical, requiring rounding to the nearest integer. On the other hand, classification offers a probability distribution over possible outcomes, making the solution more interpretable and practical.

Although the SVM method for classification used in this work shows a significant performance, it works by only considering input features frame by frame independently, without attempting to make connections between the current input and previous output(s). 
Methods based on neural networks have been found good at dealing with time-series or sequence data such as Long Short-Term Memory (LSTM) \cite{LongShortTermMemoryhochreiter1997}, Recurrent neural network (RNN) \cite{LearningPhraseRepresentationscho2014}. Furthermore, deep learning networks such as transformers have been studied for classification tasks \cite{ImageWorth16x16dosovitskiy2021}. 
For future work, the extracted features of multiple frames could be used as input as a sequence and the network can find the connection in a short period and reduce the temporal error.

\section{Conclusion}
\label{s.conclusion}

In this paper, a processing pipeline is proposed to approach track groups of people moving together and count their numbers in indoor environments. 
The pipeline is designed to handle the phenomenon of people moving independently of changes in direction or stop-and-go.
The proposed approach combines a tracker based on the extended Kalman filter framework with a classifier to obtain the number of people.
The classifier utilizes the spatial features from the range-azimuth map and Doppler frequency features with wavelet decomposition.
As a result, the pipeline outputs the location and number of people appearing over time.


The approach is tested and validated with experimental data from a 24GHz FMCW radar. The result shows that the proposed method achieves 95.59\% accuracy in counting the number of people and a combined tracking metric OSPA of 0.338.




\section*{Acknowledgments}
This research was in part financially supported by Huawei Sweden Gothenburg Research Center. The authors thank the Huawei team Zhong Chen, Yanming Wu, and Jingjing Chen for the technical discussions, and are grateful to the volunteers who participated in the data collection. 


%


\bibliographystyle{IEEEtran}
\bibliography{IEEEabrv, reference}

\begin{thebibliography}{10}
\providecommand{\url}[1]{#1}
\csname url@samestyle\endcsname
\providecommand{\newblock}{\relax}
\providecommand{\bibinfo}[2]{#2}
\providecommand{\BIBentrySTDinterwordspacing}{\spaceskip=0pt\relax}
\providecommand{\BIBentryALTinterwordstretchfactor}{4}
\providecommand{\BIBentryALTinterwordspacing}{\spaceskip=\fontdimen2\font plus
\BIBentryALTinterwordstretchfactor\fontdimen3\font minus \fontdimen4\font\relax}
\providecommand{\BIBforeignlanguage}[2]{{%
\expandafter\ifx\csname l@#1\endcsname\relax
\typeout{** WARNING: IEEEtran.bst: No hyphenation pattern has been}%
\typeout{** loaded for the language `#1'. Using the pattern for}%
\typeout{** the default language instead.}%
\else
\language=\csname l@#1\endcsname
\fi
#2}}
\providecommand{\BIBdecl}{\relax}
\BIBdecl

\bibitem{MultipleTargetPositioningyoo2019}
S.~Yoo, D.~Wang, D.-M. Seol, C.~Lee, S.~Chung, and S.~H. Cho, ``A {{Multiple Target Positioning}} and {{Tracking System Behind Brick-Concrete Walls Using Multiple Monostatic IR-UWB Radars}},'' \emph{Sensors}, vol.~19, no.~18, p. 4033, Jan. 2019.

\bibitem{PeopleTrackingandCountingReferenceDesigntexasinstruments}
\BIBentryALTinterwordspacing
{Texas Instruments}, ``{{PeopleTrackingandCounting Reference Design Using mmWave Radar Sensor}}.pdf.'' [Online]. Available: \url{https://www.ti.com/lit/ug/tidue71d/tidue71d.pdf}
\BIBentrySTDinterwordspacing

\bibitem{RadarBasedRobustPeopleninos2022}
A.~Ninos, J.~Hasch, M.~Heizmann, and T.~Zwick, ``Radar-{{Based Robust People Tracking}} and {{Consumer Applications}},'' \emph{IEEE Sensors Journal}, vol.~22, no.~4, pp. 3726--3735, Feb. 2022.

\bibitem{NoncontactExtractionBiomechanicalwang2021}
D.~Wang, J.~Park, H.-J. Kim, K.~Lee, and S.~H. Cho, ``Noncontact {{Extraction}} of {{Biomechanical Parameters}} in {{Gait Analysis Using}} a {{Multi-Input}} and {{Multi-Output Radar Sensor}},'' \emph{IEEE Access}, vol.~9, pp. 138\,496--138\,508, 2021.

\bibitem{GroupedPeopleCountingren2023}
L.~Ren, A.~Yarovoy, and F.~Fioranelli, ``Grouped {{People Counting Using}} mm-wave {{FMCW MIMO Radar}},'' \emph{IEEE Internet of Things Journal}, pp. 1--1, 2023.

\bibitem{AnalysisProcessingPipelineswang2024}
D.~Wang, F.~Fioranelli, and A.~Yarovoy, ``Analysis of {{Processing Pipelines}} for {{Indoor Human Tracking Using FMCW Radar}},'' in \emph{2024 {{IEEE Radar Conference}} ({{RadarConf24}})}, May 2024, pp. 1--6.

\bibitem{QuantitativeAssessmentPeoplewang2024}
------, ``Quantitative {{Assessment}} of {{People Tracking}} with {{FMCW MIMO Radar}},'' in \emph{2024 21st {{European Radar Conference}} ({{EuRAD}})}, Sep. 2024, pp. 380--383.

\bibitem{DensitybasedAlgorithmDiscoveringester1996}
M.~Ester, H.-P. Kriegel, J.~Sander, and X.~Xu, ``A density-based algorithm for discovering clusters in large spatial databases with noise,'' in \emph{Proceedings of the {{Second International Conference}} on {{Knowledge Discovery}} and {{Data Mining}}}, ser. {{KDD}}'96.\hskip 1em plus 0.5em minus 0.4em\relax Portland, Oregon: AAAI Press, Aug. 1996, pp. 226--231.

\bibitem{IntroductionKalmanFilterwelch1995}
G.~Welch and G.~Bishop, ``An {{Introduction}} to the {{Kalman Filter}},'' \emph{Chapel Hill, NC, USA}, 1995.

\bibitem{ReviewStatisticalDatacox1993}
I.~J. Cox, ``A review of statistical data association techniques for motion correspondence,'' \emph{International Journal of Computer Vision}, vol.~10, no.~1, pp. 53--66, Feb. 1993.

\bibitem{ProbabilisticDataAssociation2009a}
``The probabilistic data association filter,'' \emph{IEEE Control Systems}, vol.~29, no.~6, pp. 82--100, Dec. 2009.

\bibitem{MaximalOverlapDiscretexiao2020}
F.~Xiao, T.~Lu, M.~Wu, and Q.~Ai, ``Maximal overlap discrete wavelet transform and deep learning for robust denoising and detection of power quality disturbance,'' \emph{IET Generation, Transmission \& Distribution}, vol.~14, no.~1, pp. 140--147, 2020.

\bibitem{MaximalOverlapDiscretequilty2021}
J.~Quilty and J.~Adamowski, ``A maximal overlap discrete wavelet packet transform integrated approach for rainfall forecasting -- {{A}} case study in the {{Awash River Basin}} ({{Ethiopia}}),'' \emph{Environmental Modelling \& Software}, vol. 144, p. 105119, Oct. 2021.

\bibitem{ChangeDetectionTimealarcon-aquino2009}
\BIBentryALTinterwordspacing
V.~{Alarcon-Aquino} and J.~A. Barria, ``Change detection in time series using the maximal overlap discrete wavelet transform,'' \emph{Latin American applied research}, vol.~39, no.~2, pp. 145--152, Jun. 2009. [Online]. Available: \url{https://www.scielo.org.ar/scielo.php?script=sci_abstract&pid=S0327-07932009000200009&lng=es&nrm=iso&tlng=en}
\BIBentrySTDinterwordspacing

\bibitem{KNNModelBasedApproachguo2003}
G.~Guo, H.~Wang, D.~Bell, Y.~Bi, and K.~Greer, ``{{KNN Model-Based Approach}} in {{Classification}},'' in \emph{On {{The Move}} to {{Meaningful Internet Systems}} 2003: {{CoopIS}}, {{DOA}}, and {{ODBASE}}}, R.~Meersman, Z.~Tari, and D.~C. Schmidt, Eds.\hskip 1em plus 0.5em minus 0.4em\relax Berlin, Heidelberg: Springer, 2003, pp. 986--996.

\bibitem{EmpiricalStudyNaiverisha}
I.~Rish, ``An empirical study of the naive {{Bayes}} classifier.''

\bibitem{SupportVectorMachineshearst1998}
M.~Hearst, S.~Dumais, E.~Osuna, J.~Platt, and B.~Scholkopf, ``Support vector machines,'' \emph{IEEE Intelligent Systems and their Applications}, vol.~13, no.~4, pp. 18--28, Jul. 1998.

\bibitem{RandomForestRemotebelgiu2016}
M.~Belgiu and L.~Dr{\u a}gu{\c t}, ``Random forest in remote sensing: {{A}} review of applications and future directions,'' \emph{ISPRS Journal of Photogrammetry and Remote Sensing}, vol. 114, pp. 24--31, Apr. 2016.

\bibitem{AzureKinectDK}
\BIBentryALTinterwordspacing
``Azure {{Kinect DK}} -- {{Develop AI Models}} {\textbar} {{Microsoft Azure}}.'' [Online]. Available: \url{https://azure.microsoft.com/en-us/products/kinect-dk}
\BIBentrySTDinterwordspacing

\bibitem{Detectron2yuxinwuandalexanderkirillovandfranciscomassaand2025}
\BIBentryALTinterwordspacing
{Yuxin Wu and Alexander Kirillov and Francisco Massa and} and {Wan-Yen Lo and Ross Girshick}, ``Detectron2,'' Meta Research, Feb. 2025. [Online]. Available: \url{https://github.com/facebookresearch/detectron2}
\BIBentrySTDinterwordspacing

\bibitem{ConsistentMetricPerformanceschuhmacher2008}
D.~Schuhmacher, B.-T. Vo, and B.-N. Vo, ``A {{Consistent Metric}} for {{Performance Evaluation}} of {{Multi-Object Filters}},'' \emph{IEEE Transactions on Signal Processing}, vol.~56, no.~8, pp. 3447--3457, Aug. 2008.

\bibitem{LongShortTermMemoryhochreiter1997}
S.~Hochreiter and J.~Schmidhuber, ``Long {{Short-Term Memory}},'' \emph{Neural Computation}, vol.~9, no.~8, pp. 1735--1780, Nov. 1997.

\bibitem{LearningPhraseRepresentationscho2014}
K.~Cho, B.~van Merrienboer, C.~Gulcehre, D.~Bahdanau, F.~Bougares, H.~Schwenk, and Y.~Bengio, ``Learning {{Phrase Representations}} using {{RNN Encoder-Decoder}} for {{Statistical Machine Translation}},'' Sep. 2014.

\bibitem{ImageWorth16x16dosovitskiy2021}
A.~Dosovitskiy, L.~Beyer, A.~Kolesnikov, D.~Weissenborn, X.~Zhai, T.~Unterthiner, M.~Dehghani, M.~Minderer, G.~Heigold, S.~Gelly, J.~Uszkoreit, and N.~Houlsby, ``An {{Image}} is {{Worth}} 16x16 {{Words}}: {{Transformers}} for {{Image Recognition}} at {{Scale}},'' Jun. 2021.

\end{thebibliography}

\vspace{-33pt}
\begin{IEEEbiography}[{\includegraphics[width=1in,height=1.25in,clip,keepaspectratio]{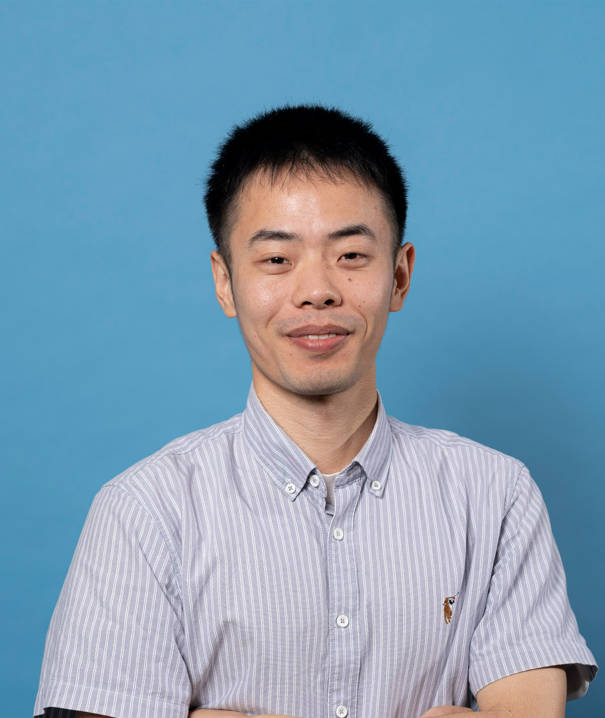}}]{Dingyang Wang} (Member, IEEE) received a B.S. degree in electronic engineering from Andong National University, Gyeongsangbuk-do, South Korea, in 2015.
He joined the Radar Computing Lab with a joint M.S. and Ph.D. program in March 2016. 
In Feb 2022, he received his Ph.D. on gait analysis with MIMO radar under the supervision of Prof. SUNG HO CHO at Hanyang University, Seoul, South Korea.
Since April 2023, he became a postdoctoral researcher at the MS3 (Microwave Sensing, Signals and Systems) group, Faculty of Electrical Engineering, Delft University of Technology.
His research includes radar signal processing and multi-target tracking.
\end{IEEEbiography}

\begin{IEEEbiography}[{\includegraphics[width=1in,height=1.25in,clip,keepaspectratio]{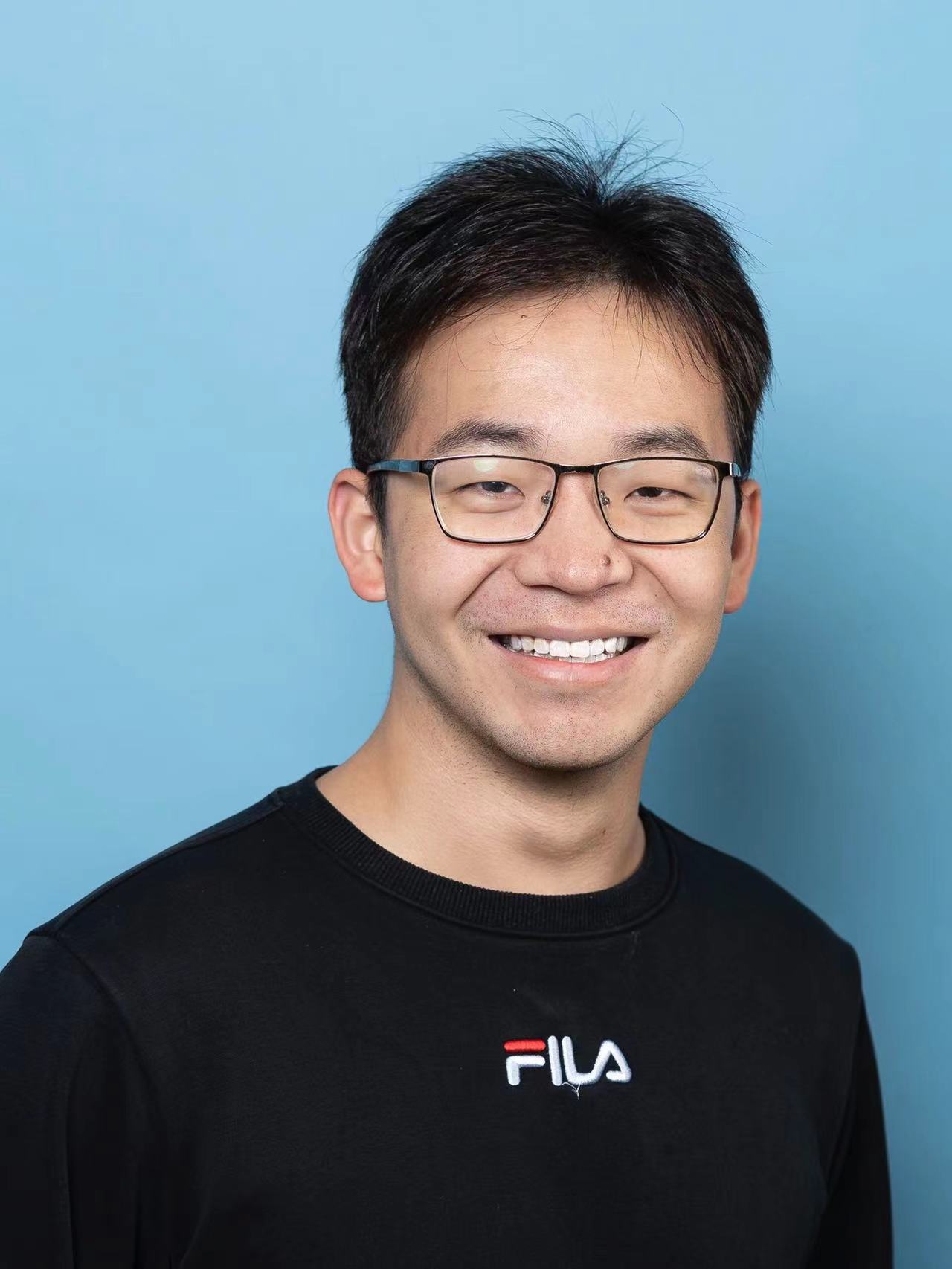}}]{Sen Yuan} (Member, IEEE) was born in Shanxi province, China, in 1998. He received his Ph.D. degree from the Delft University of Technology, Delft, The Netherlands, in 2024, in electrical engineering. He is currently a Post-Doctoral Researcher with the Group of Microwave Sensing, Signals and Systems (MS3), Delft University of Technology. His research interests include SAR imaging, signal processing for radar system, and new scheme of radar system design. He was the recipient of the European Microwave Association Student Grant in 2021, 2022, 2023 and 2024. He is currently an associate member of the IEEE Signal Processing Society Autonomous Systems Initiative (ASI), and he has served as a reviewer of many IEEE journals.
\end{IEEEbiography}

\begin{IEEEbiography}[{\includegraphics[width=1in,height=1.25in,clip,keepaspectratio]{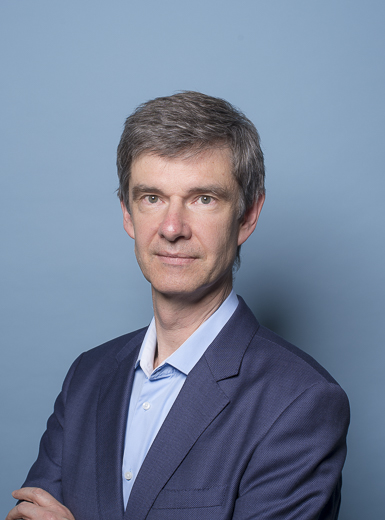}}]{Alexander Yarovoy} (Fellow, IEEE) received the Diploma degree (Hons.) in radiophysics and elec- tronics and the Candidate Physics and Mathematical Sciences and the Doctor Physics and Mathemati- cal Sciences degrees in radiophysics from Kharkov State University, Ukraine, in 1984, 1987, and 1994, respectively. In 1987, he joined the Department of Radiophysics, Kharkov State University, as a Researcher, where he became a Full Professor in 1997. From September 1994 to September 1996, he was with the Technical University of Ilmenau, Germany, as a Visiting Researcher. Since 1999, he has been with the Delft Uni- versity of Technology, The Netherlands, where he has been leading the Chair of Microwave Sensing, Systems and Signals, since 2009. He has authored and coauthored more than 500 scientific or technical papers, seven patents, and 14 book chapters. His current research interests include high-resolution radar, microwave imaging, and applied electromagnetics (in particular, UWB antennas). He was a recipient of the European Microwave Week Radar Award for the paper that best advances the state-of-the-art in radar technology in 2001 (together with L.P. Ligthart and P. van Genderen) and in 2012 (together with T. Savelyev). In 2010 together with D. Caratelli, he got the Best Paper Award of the Applied Computational Electromagnetic Society (ACES). He served as the General TPC Chair for the 2020 European Microwave Week (EuMW- 20), the Chair and the TPC Chair for the Fifth European Radar Conference (EuRAD-08), and the Secretary for the First European Radar Conference (EuRAD-04). He also served as the Co-Chair and the TPC Chair for the Tenth International Conference on GPR (GPR2004). He serves as an Associate Editor for IEEE TRANSACTION ON RADAR SYSTEMS. From 2011 to 2018, he served as an Associate Editor for the International Journal of Microwave and Wireless Technologies. From 2008 to 2017, he served as the Director for the European Microwave Association (EuMA).
\end{IEEEbiography}

\begin{IEEEbiography}[{\includegraphics[width=1in,height=1.25in,clip,keepaspectratio]{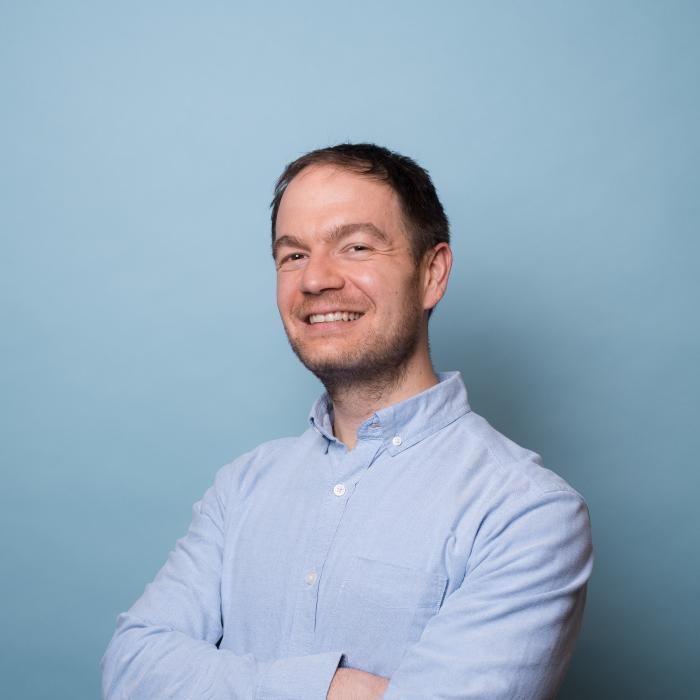}}]{Francesco Fioranelli} (Senior Member, IEEE) 
received his Laurea (BEng, cum laude) and Laurea Specialistica (MEng, cum laude) degrees in telecommunication engineering from the Università Politecnica delle Marche, Ancona, Italy, in 2007 and 2010, respectively, and the Ph.D. degree from Durham University, U.K., in 2014. He is currently an Associate Professor at TU Delft in the Netherlands, and was an Assistant Professor at the University of Glasgow (2016-2019) and Research Associate at University College London (2014-2016). 
	
His research interests include the development of radar systems and automatic classification for human signatures analysis in healthcare and security, drones and UAVs detection and classification, automotive radar, wind farm and sea clutter. He has authored over 190 peer-reviewed publications, edited the books on “Micro-Doppler Radar and Its Applications” and "Radar Countermeasures for Unmanned Aerial Vehicles" published by IET-Scitech in 2020, and received four best paper awards and the IEEE AESS Fred Nathanson Memorial Radar Award 2024.

\end{IEEEbiography}

\vfill

\end{document}